\documentclass[%
reprint,
superscriptaddress,
amsmath,amssymb,
aps,
prc,
]{revtex4-1}

\usepackage{graphicx}
\usepackage{dcolumn}
\usepackage{bm}

\usepackage{amsmath}
\usepackage{color}
\usepackage[utf8]{inputenc}
\usepackage{multirow}

\begin{document}

\title{Structure of $^{9}$C through proton resonance scattering with Texas Active Target detector}

\author{J. Hooker}
\altaffiliation[Present address: ]{Department of Physics \& Astronomy, University of Tennessee, Knoxville, TN 37996, USA}
\affiliation{%
Department of Physics \& Astronomy, Texas A\&M University, College Station, TX 77843, USA
}
\affiliation{%
Cyclotron Institute, Texas A\&M University, College Station, TX 77843, USA
}
\author{G.~V. Rogachev}
\email{rogachev@tamu.edu}
\affiliation{%
Department of Physics \& Astronomy, Texas A\&M University, College Station, TX 77843, USA
}
\affiliation{%
Cyclotron Institute, Texas A\&M University, College Station, TX 77843, USA
}
\affiliation{%
Nuclear Solutions Institute, Texas A\&M University, College Station, TX 77843, USA
}
\author{E. Koshchiy}
\affiliation{%
Cyclotron Institute, Texas A\&M University, College Station, TX 77843, USA
}
\author{S. Ahn}
\affiliation{%
Cyclotron Institute, Texas A\&M University, College Station, TX 77843, USA
}
\author{M. Barbui}
\affiliation{%
Cyclotron Institute, Texas A\&M University, College Station, TX 77843, USA
}
\author{V.~Z. Goldberg}
\affiliation{%
Cyclotron Institute, Texas A\&M University, College Station, TX 77843, USA
}
\author{C. Hunt}
\affiliation{%
Department of Physics \& Astronomy, Texas A\&M University, College Station, TX 77843, USA
}
\affiliation{%
Cyclotron Institute, Texas A\&M University, College Station, TX 77843, USA
}
\author{H. Jayatissa}
\altaffiliation[Present address: ]{Physics Division, Argonne National Laboratory, Argonne, IL 60439, USA}
\affiliation{%
Department of Physics \& Astronomy, Texas A\&M University, College Station, TX 77843, USA
}
\affiliation{%
Cyclotron Institute, Texas A\&M University, College Station, TX 77843, USA
}
\author{E.C.~Pollacco}
\affiliation{IRFU, CEA, Saclay, Gif-Sur-Ivette, France}
\author{B.~T. Roeder}
\affiliation{%
Cyclotron Institute, Texas A\&M University, College Station, TX 77843, USA
}
\author{A. Saastamoinen}
\affiliation{%
Cyclotron Institute, Texas A\&M University, College Station, TX 77843, USA
}
\author{S. Upadhyayula}
\affiliation{%
Department of Physics \& Astronomy, Texas A\&M University, College Station, TX 77843, USA
}
\affiliation{%
Cyclotron Institute, Texas A\&M University, College Station, TX 77843, USA
}

\date{\today}

\begin{abstract}
\begin{description}
\item[Background]
Level structure of the most neutron deficient nucleon-bound carbon isotope, $^9$C, is not well known. Definitive spin-parity assignments are only available for two excited states. No positive parity states have been conclusively identified so far and the location of the sd-shell in A=9 T=3/2 isospin quadruplet is not known.
\item[Purpose]
We have studied the level structure of exotic nucleus $^9$C at excitation energies below 6.4 MeV.
\item[Methods]
Excited states in $^9$C were populated in $^8$B+p resonance elastic scattering and excitation functions were measured using active target approach.
\item[Results] Two excited states in $^9$C were conclusively observed, and R-matrix analysis of the excitation functions was performed to make the spin-parity assignments. The first positive parity state in A=9 T=3/2 nuclear system, the 5/2$^+$ resonance at 4.3 MeV, has been identified.
\item[Conclusions] The new 5/2$^+$ state at 4.3 MeV in $^9$C is a single-particle $\ell=0$ broad resonance and it determines the energy of the 2s shell. The 2s shell in this exotic nucleus appears well within the region dominated by the p-shell states.
\end{description}
\end{abstract}

\pacs{Valid PACS appear here}
\maketitle


\section{\label{sec:Introduction}Introduction}

Enormous progress has been achieved over the past two decades in describing the properties of light nuclei starting from interacting nucleons and using realistic two-nucleon and three-nucleon forces or chiral interactions. Sophisticated methods, such as Quantum Monte Carlo approach \cite{Pieper2002}, No Core Shell Model \cite{Navratil1998},  No Core Configuration Interaction \cite{Maris2009}, coupled-cluster theory \cite{Hagen2007} have been developed to make robust predictions of ground-state energies \cite{Binder2018}, level structure \cite{Pieper2002}, spectroscopic factors and partial widths \cite{Nollett2012}, scattering phase shifts \cite{Forssen2011}, electromagnetic moments and transitions \cite{Pastore2013} in light nuclei. Reliable experimental benchmarks are necessary to facilitate further theoretical progress. The focus of this experimental study is $^9$C. The first {\it ab initio} calculations for A=9 T=3/2 systems have become available in 1998 \cite{Navratil1998}. Only few levels were experimentally known at the time, and robust spin-parity assignments were available only for two of them. Therefore, a detailed comparison to the experimental data was handicapped. Few experiments on $^9$Li and $^9$C have been performed since then, and more experimental information has become available. Yet, the spin-parity has been reliably established for only one more state. Moreover, all known states are negative parity (p-shell) states and no evidence for positive parity states have been observed so far, making the energy of the sd-shell in this system an open question. The goal of this study was to improve our knowledge of the level structure of $^9$C in general and locate the onset of the 2s-shell in this A=9 T=3/2 system in particular. The 2s shell plays an important role in the structure and stability of light exotic nuclei. The 2s ground state in $^{11}$Be is now a textbook case, and it is well recognized that the 2s shell dominated the structure of ground states of some exotic nuclei, such as $^9$He  \cite{Uberseder2016}, $^{10}$N \cite{Hooker2017}, $^{11}$N \cite{Axelsson1996} and $^{14}$F \cite{Goldberg2010}.

In addition, this experimental study was the commissioning run of the Texas Active Target detector system (TexAT) built for experiments with rare isotope beams at Texas A\&M University Cyclotron Institute and elsewhere. The current state of experimental knowledge of $^9$C spectroscopy is reviewed below.

$^{9}$C is a proton drip-line carbon isotope that has the largest Z/N ratio (2) among all nucleon-bound nuclei in the nuclear chart (same as $^3$He). It has a half-life of 126.5 ms, and proton separation energy of 1.3 MeV \cite{Tilley2004}. The ground state of $^{9}$C was discovered in 1964 by Cerny \emph{et al.} \cite{Cerny:1964}, the first excited state was observed in 1974 at 2.2 MeV \cite{Benenson:1974}, and an excited state at 3.3 MeV was reported in \cite{Golovkov:1991}, but it was not confirmed in later experiments. All of these studies used the $^{12}$C($^3$He,$^6$He) reaction. More recently, the $5/2^{-}$ state at 3.6 MeV was observed in the excitation function for $^8$B+p elastic scattering, which was measured up to 4.5 MeV excitation energy at just one angle \cite{Rogachev:2006}. Finally, the structure of $^{9}$C was studied using inelastic scattering of a $^{9}$C beam on a $^{9}$Be target \cite{Brown:2017}. Two states were observed by measuring $^{8}$B and protons in coincidence, corresponding to the first and second excited states at 2.2 and 3.55 MeV, in good agreement with Refs. \cite{Benenson:1974,Rogachev:2006}. The $^{7}$Be+2p decay channel was also measured and the authors claim to observe two resonances at 4.40 MeV and 5.69 MeV \cite{Brown:2017}.

We performed the study of $^{9}$C using $^{8}$B+p resonance elastic scattering to extend the excitation function to higher energies compared to the previous measurement \cite{Rogachev:2006}, and most importantly to obtain the angular distribution. The latter allowed us to conclusively identify the first positive parity state in $^9$C.

\section{\label{sec:Experiment} Experiment and Analysis}

The $^{8}$B beam was produced in the $^{6}$Li($^{3}$He,n)$^8$B reaction using the Momentum Achromat Recoil Spectrometer (MARS) at the Cyclotron Institute at Texas A\&M University \cite{Tribble:1989}. The primary $^6$Li beam of energy 13.2 MeV/u from K150 cyclotron was directed to the LN cooled, 9.2 cm long, gas cell with 4 $\mu$m thick and 19 mm diameter Havar entrance and exit windows. Pressure of the $^3$He gas inside the cell was 810 Torr. The resulting $^{8}$B beam had an energy of 60.8 MeV with an intensity of $10^{3}$ pps, energy spread of 1.2 MeV and a beam spot size of about 10 mm FWHM. The main contaminant was the scattered ions of  primary $^{6}$Li beam at $\sim$1.6 \% level.

\begin{figure}
\includegraphics[width=1.0\linewidth]{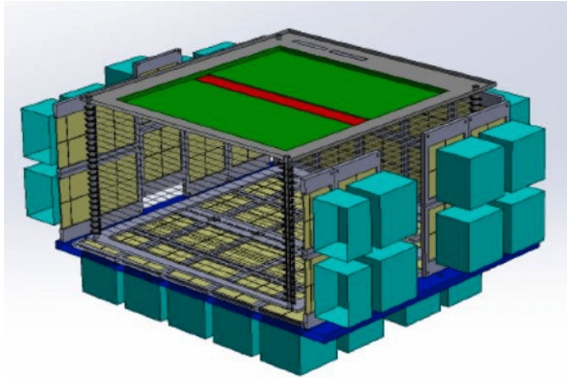}
\caption{(Color online) TexAT Assembly with one side removed. The top part is the Micromegas where the red portion depicts the central pads and the green are the side regions. The Si detectors (yellow) are each backed by a CsI (turquoise) detector. The beam travels from right to left along the central pads \cite{Koshchiy:2015}. \label{fig:texat}}
\end{figure}

A brief overview of the TexAT detector, shown in Fig.\ \ref{fig:texat}, is provided below. Technical details on TexAT detector system, the TexAT GEANT MonteCarlo simulation, and the TexAT 3D track reconstruction procedures are described in Ref. \cite{TexAT2019}.

TexAT is a time projection chamber (TPC) with a planar geometry. It is based on highly segmented Micromegas \cite{Giomataris:1995} detector which provides particle ID and 3D tracking for the incoming beam ions and charged products of nuclear reactions. The Micromegas detector has an active area of 224x224 mm$^2$ and consists of 1024 channels, of which 768 are in the central (beam) region of rectangular 3.5$\times$1.75 mm$^2$ pads arranged into 6 columns and 128 rows along the beam axis. Pads in the ``sides'' region of the Micromegas detector are multiplexed into chains and strips, running parallel and perpendicular to the beam axis respectively, for a total of 64 chains and 64 strips per each side. The multiplexing is used to reduce the channel count. In the central region the 3D image of the tracks is produced using individual pads and ionization electrons drift times, while in the side regions the drift times are also used to match chains and strips. Further track recognition is performed using the Hough transform \cite{Duda:1972}, which allows reliable identification of tracks even in sub-optimal noise conditions. In addition to the TPC, TexAT includes a windowless ionization chamber (IC) located near the scattering chamber entrance window, and an array of silicon detectors, backed by CsI(Tl) scintillators, that surround the TPC on all but the Micromegas sides. IC is used for incoming beam ions particle ID and overall normalization. Normally, a total of 50 5$\times$5 cm$^2$ silicon detectors (700-1000 $\mu$m thick and each consisting of four 2.5$\times$2.5 cm$^{2}$ square segments), backed by 50 5$\times$5$\times$4 cm$^3$ CsI(Tl) scintillator detectors, read out by Si pin-diodes, are used by TexAT. However, only 9 Si+CsI(Tl) pairs were installed in the most forward region of TexAT for the commissioning run. General Electronics for TPCs (GET) \cite{get} is used for all TexAT channels. The data is recorded using Narval DAQ \cite{Narval}.

\begin{figure}
    \includegraphics[width=0.5\textwidth]{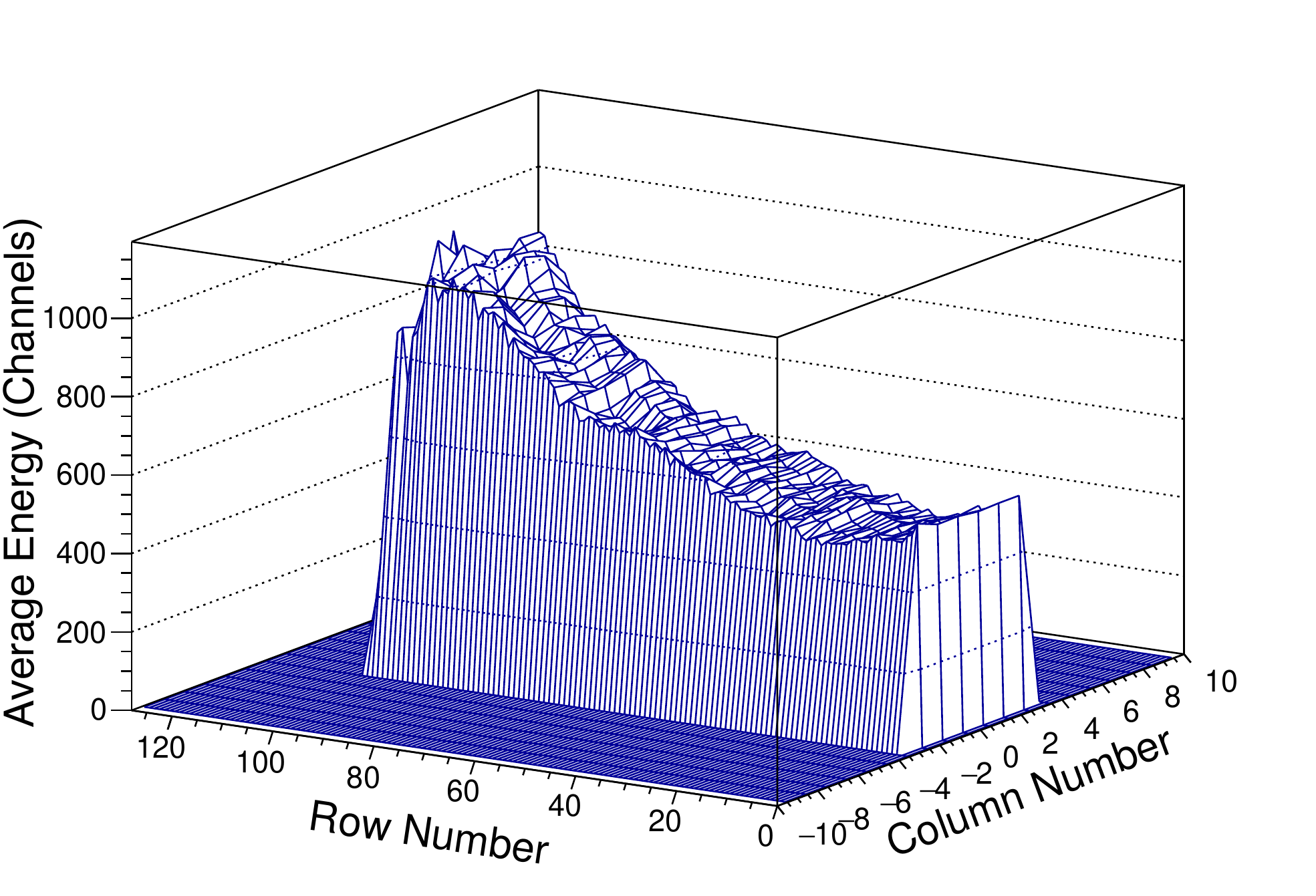}
    \caption{The energy deposited in each of the central region pads. The Bragg peak occurs around row number 90.   \label{fig:9CBeamEnergy}}
\end{figure}

The $^8$B beam enters the TexAT scattering chamber through a 4 $\mu$m thick Havar window. The target Methane gas (CH$_{4}$, research grade 99.999\% purity) pressure was adjusted to stop the incoming $^8$B beam ions before the last 16 rows of the central pads - 435 Torr. The profile of average energy losses of the beam ions, shown in Fig.\ \ref{fig:9CBeamEnergy}, reflects the location of the Bragg peak and the range of $^8$B ions. Anode biasing scheme is utilized in TexAT for the Micromegas detector, with specific sets of pads biased individually.


This arrangement makes it possible to apply different voltages for different regions of the detector, and as a result to have different gas gains. Hence, we used low gas gain (400 V tension) in the first 7/8 of the central pads region to record tracks of the beam ions, and high gas gain (600 V tension) in the last 1/8 - the furthermost 16 rows from the Havar entrance window. This allowed us to record tracks of light (proton) recoils in the last Micromegas section. The side regions all had high (570 V) tension for high gas gain.

\begin{figure}
\centering
\includegraphics[scale=0.45]{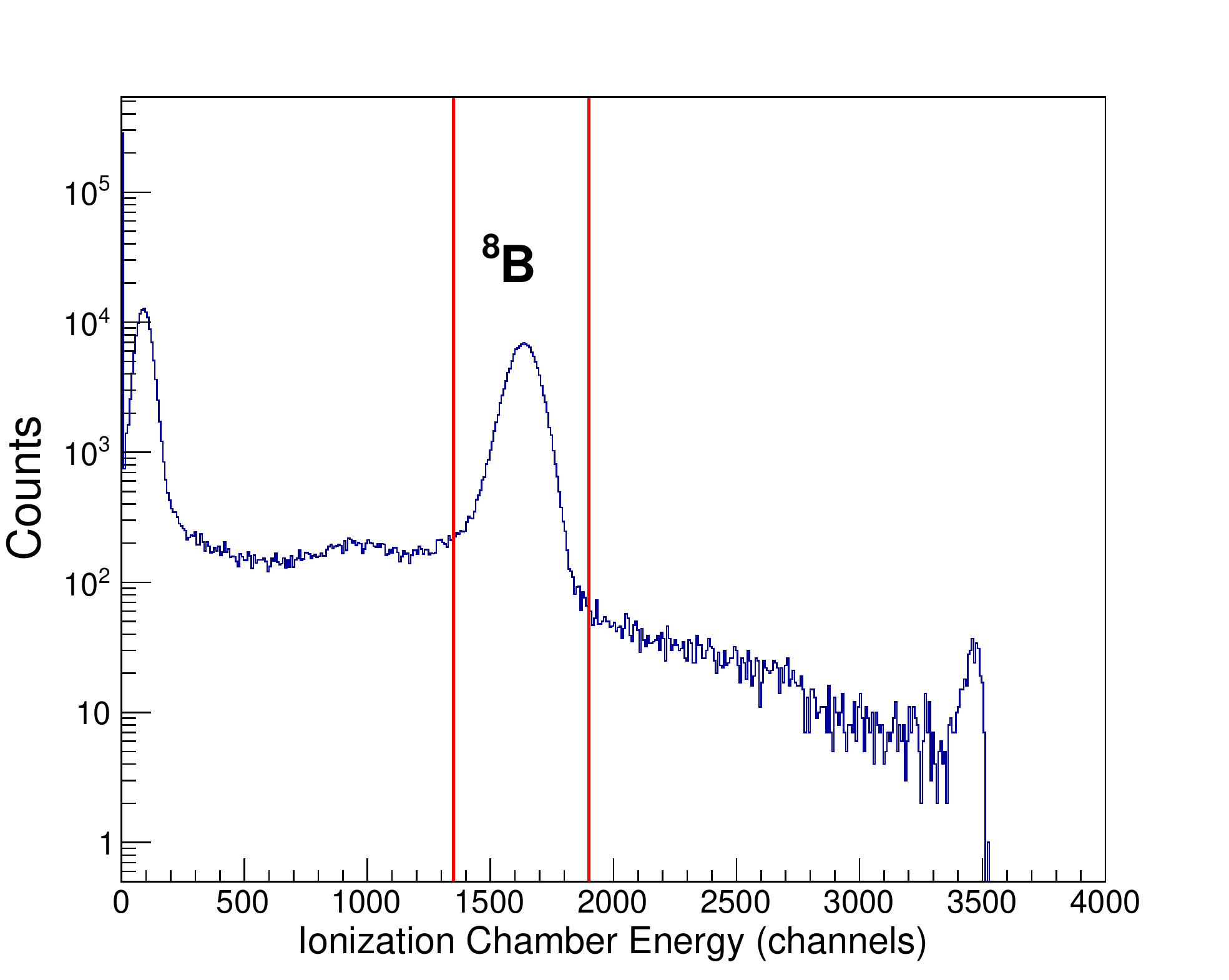}
\caption{Energy deposition in the ionization chamber at the entrance of TexAT. A peak at channel 1700 corresponds to the $^8$B beam ions. The red lines show the energy gate that was applied to select events associated with $^8$B.  \label{fig:icE}}
\end{figure}

\begin{figure}
\centering
\includegraphics[scale=0.45]{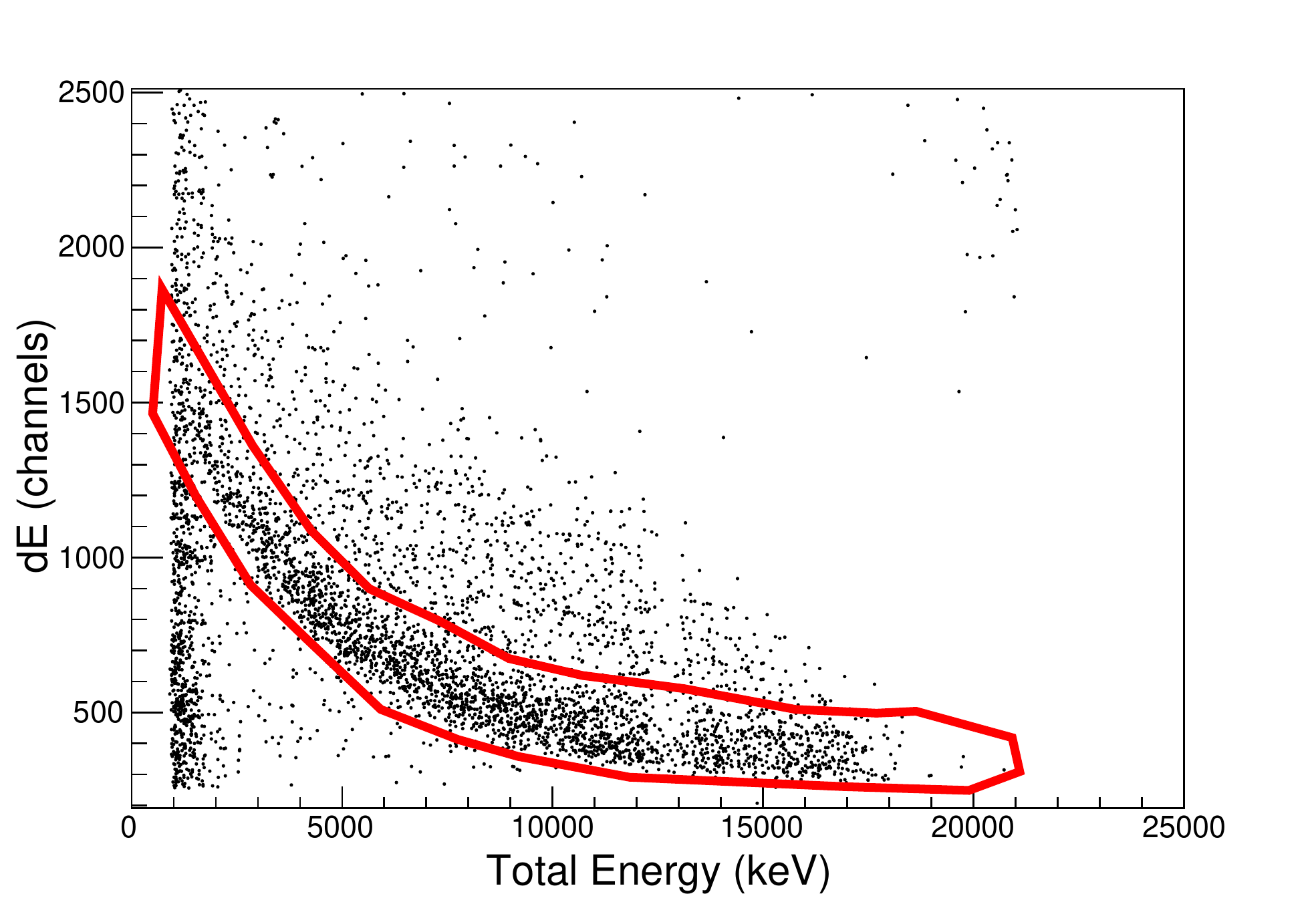}
\caption{(Color online) Scatter plot of the specific energy loss per unit pad in the Micromegas detector ($\Delta$E/pad) versus the sum of energies deposited in the Si and CsI(Tl) detectors for detectors with a c.m.\ angle of 100-145$^{\circ}$. The red band shows the ``proton'' cut. \label{fig:dETotalE}}
\end{figure}

\begin{figure}
\centering
\includegraphics[scale=0.45]{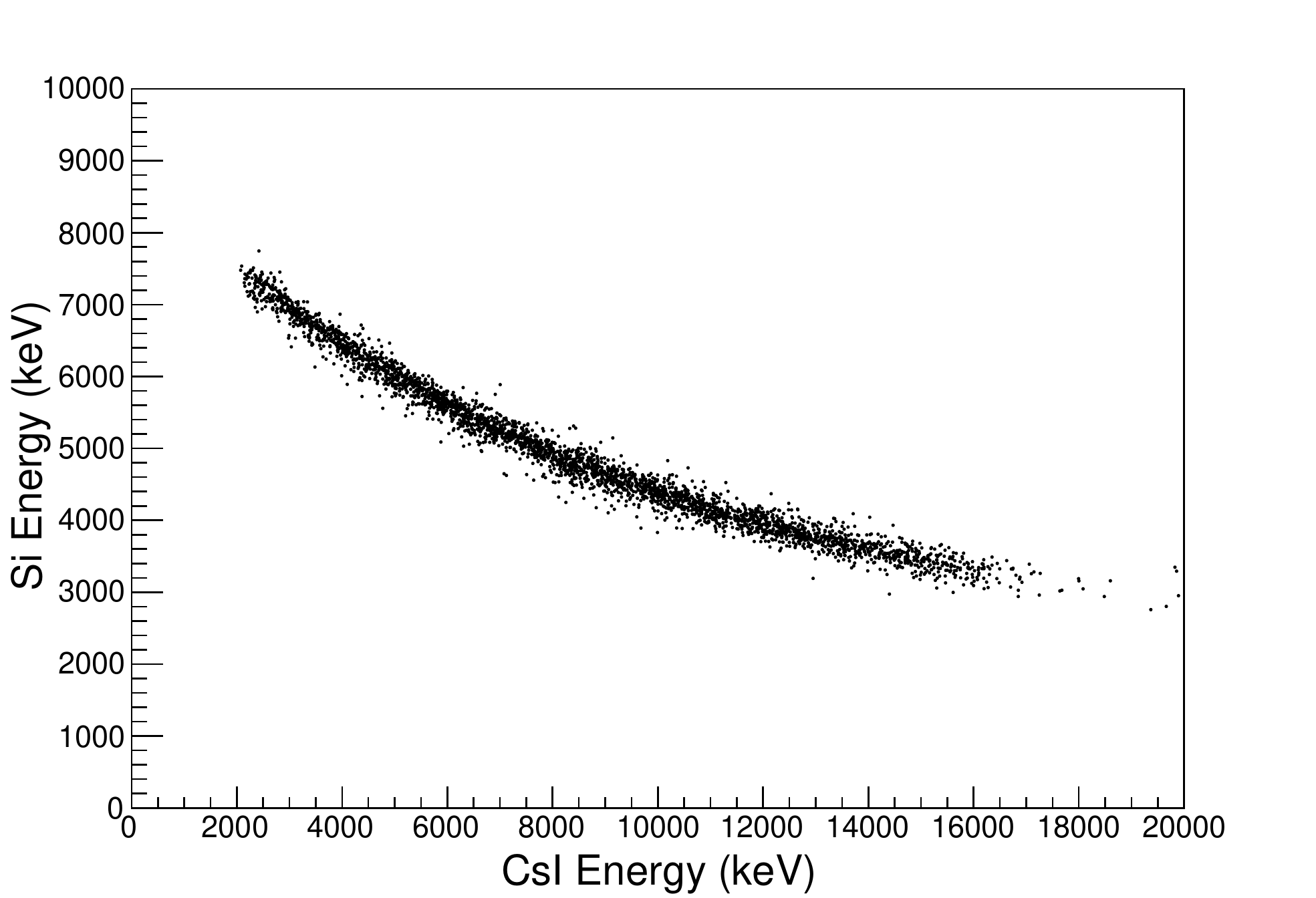}
\caption{Energy in the CsI plotted versus energy deposited in the Si detector for proton events that punch-through the Si detectors and above the threshold in the CsI for detectors with a c.m.\ angle 155-175$^{\circ}$. \label{fig:siEcsiE}}
\end{figure}

The event ID for $^8$B+p elastic scattering with TexAT is robust. We first gate on the energy deposition in the ionization chamber to select the events associated with $^8$B (see Fig.\ \ref{fig:icE}). We then identify proton events using the $\Delta$E-E scatter plot of specific energy loss per unit pad in the Micromegas detector vs total energy measured in Si+CsI(Tl) (Fig.\ \ref{fig:dETotalE}). For higher energy protons that punch through the silicon detectors, we apply an additional cut in the $\Delta$E-E scatter plot of energy deposition in silicon detector vs energy in CsI(Tl) (Fig.\ \ref{fig:siEcsiE}). This additional step was not strictly necessary but we used it to check that the first $\Delta$E-E selection produced a clean proton event identification. The kinetic energy of protons at the vertex location was determined as the sum of the measured energies in the Si and CsI(Tl) detectors and the calculated proton energy loss in the gas on event by event basis. For the latter we used code SRIM \cite{Srim:2010} and the measured reaction vertex location (see below). For the reference, energy loss in the gas for 8.7 MeV protons that correspond to zero degrees $^8$B+p elastic scattering events at 2.45 MeV in c.m. is $\sim$600 keV on average.

\begin{figure}
\centering
\includegraphics[scale=0.45]{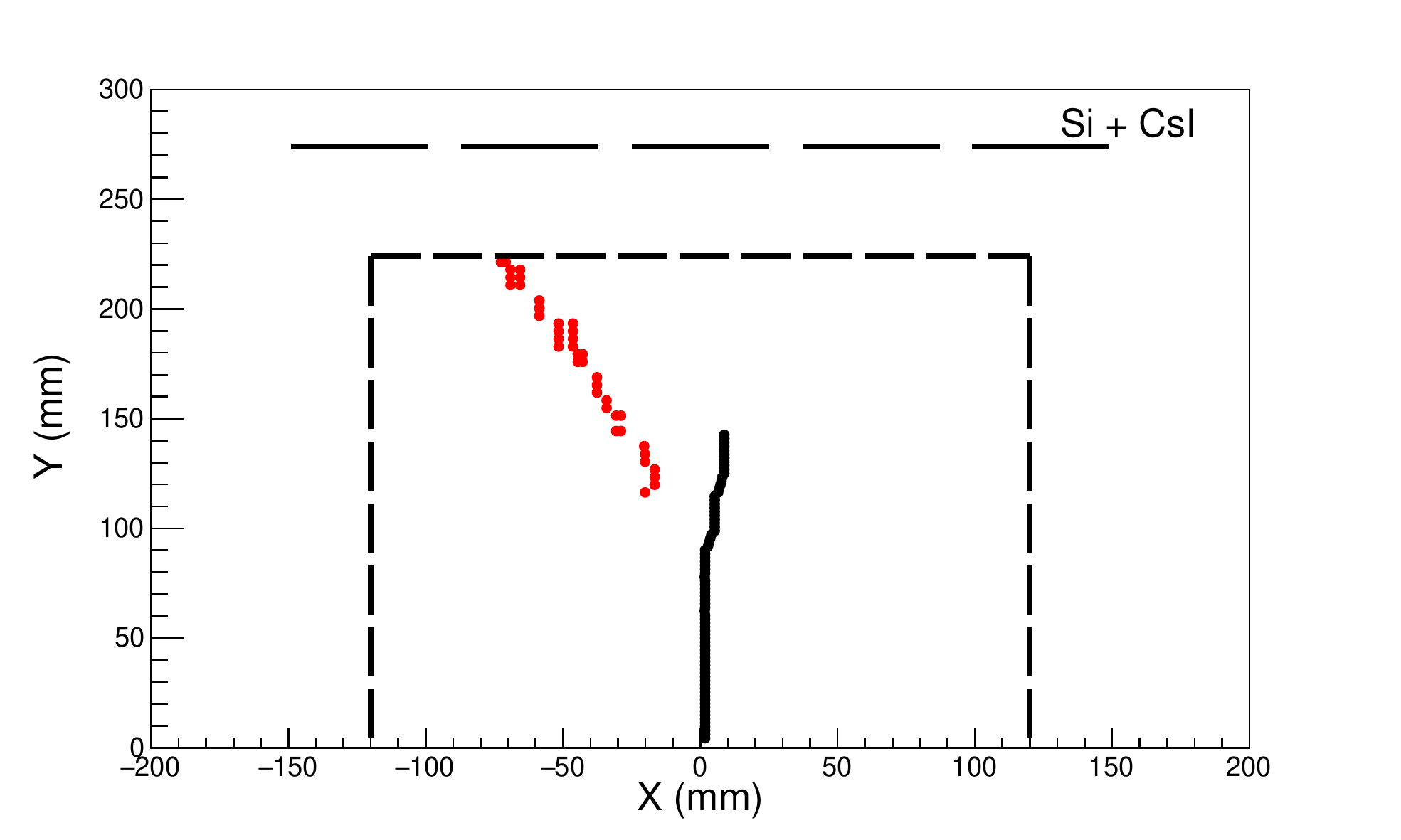}
\caption{(Color online) \label{fig:sideprotontrack} A 2D projection onto the micromegas plane for a typical $^8$B+p elastic scattering event. The $^8$B projectile and recoil tracks are shown in black. The proton track, produced by matching strips and chains in the multiplexed high-gain side region of the micromegas detector, is shown in red. Active area of the micromegas detector is shown by the dashed lines and the location of the Si+CsI(Tl) telescopes is shown with bold solid lines. Note that proton track is not visible until it gets to the high gain region of the micromegas, this is why the proton track does not start at the reaction vertex location.}
\end{figure}

We then used 3D tracking in the TPC to reconstruct the complete kinematics of the event. The details of the tracking
procedure are described in Ref. \cite{TexAT2019}. Briefly, the reaction vertex location for the proton tracks in the side
region is determined using fitting of the three tracks (beam ion track, proton recoil track in the side region, and heavy ion
recoil track) with straight lines. A typical event that has a proton scattered into the side region of the micromegas detector is
shown in Fig.\ \ref{fig:sideprotontrack}. If the reaction vertex is outside of the active area of
the TPC then only the proton track was used to reconstruct the vertex location, in which case the vertex location was
determined as a crossover point between a proton track (projected onto the plane of the micromegas detector) and the
beam axis. Vertex location reconstructed this way is plotted against the sum of the energies measured by the Si and CsI(Tl)
detectors in the bottom panel of Fig. \ref{fig:vertexTotalEnergy}. Negative values for the reaction vertex location correspond
to reaction events that occurred before the active region of micromegas detector, but produced light recoils that punched
through the gas and hit the Si array.
For the events with proton tracks that appear only in the central region (small lab. scattering angles) the vertex location was
reconstructed using the location of the Bragg peak for the heavy recoils. We have chosen this approach because it allows
for an extension of the vertex location reconstruction into the region outside of the TPC's active area. Vertex location
reconstructed using the Bragg peak is plotted against the energy in the Si and CsI(Tl) detectors in the top panel of Fig.
\ref{fig:vertexTotalEnergy}. The downside of this approach is that it does not work for the higher energy events which
produce heavy recoils too far to reach the active region of the TPC. This is why the top panel of Fig.
\ref{fig:vertexTotalEnergy} does not extend as far back as the bottom panel, that does not suffer from this limitation.
Elastic scattering kinematics results in a well-defined trend of vertex location vs total energy of light recoil, and it was used
for additional selection of elastic scattering events. Histogram of reconstructed vertex locations for
c.m.\ energies of $2.4-2.5$ MeV is shown in Fig.\ \ref{fig:vertexCMEnergy24_25}. An obvious peak between 55 and 90 mm
(measured from the start of the micromegas active region) corresponds to the elastic scattering events.

Clearly visible gaps in the spectrum of protons in Figs. \ref{fig:dETotalE} and \ref{fig:vertexTotalEnergy} are due to the threshold of the CsI(Tl) detectors. These gaps occur in the energy ranges 9-9.5 MeV and 12.5-13 MeV for the 700 $\mu$m and the 1000 $\mu$m thick detectors respectively. As no narrow resonances have been observed or are expected in the measured excitation energy region in $^{9}$C, and to avoid discontinuities in the excitation function, the events with energies 0.5 MeV below the observed gaps were randomly sampled and energy was added back to some of them. This sampling was guided by the Monte Carlo simulation of the punch-through events, taking into account the CsI(Tl) detectors threshold. The c.m.\ energy regions affected by this procedure are 2.4-2.7 MeV for the larger c.m.\ scattering angles (Fig.\ \ref{fig:EF}a) and 4.1-4.4 MeV for the smaller scattering angles (Fig.\ \ref{fig:EF}b).

\begin{figure}
\centering
\includegraphics[scale=0.45]{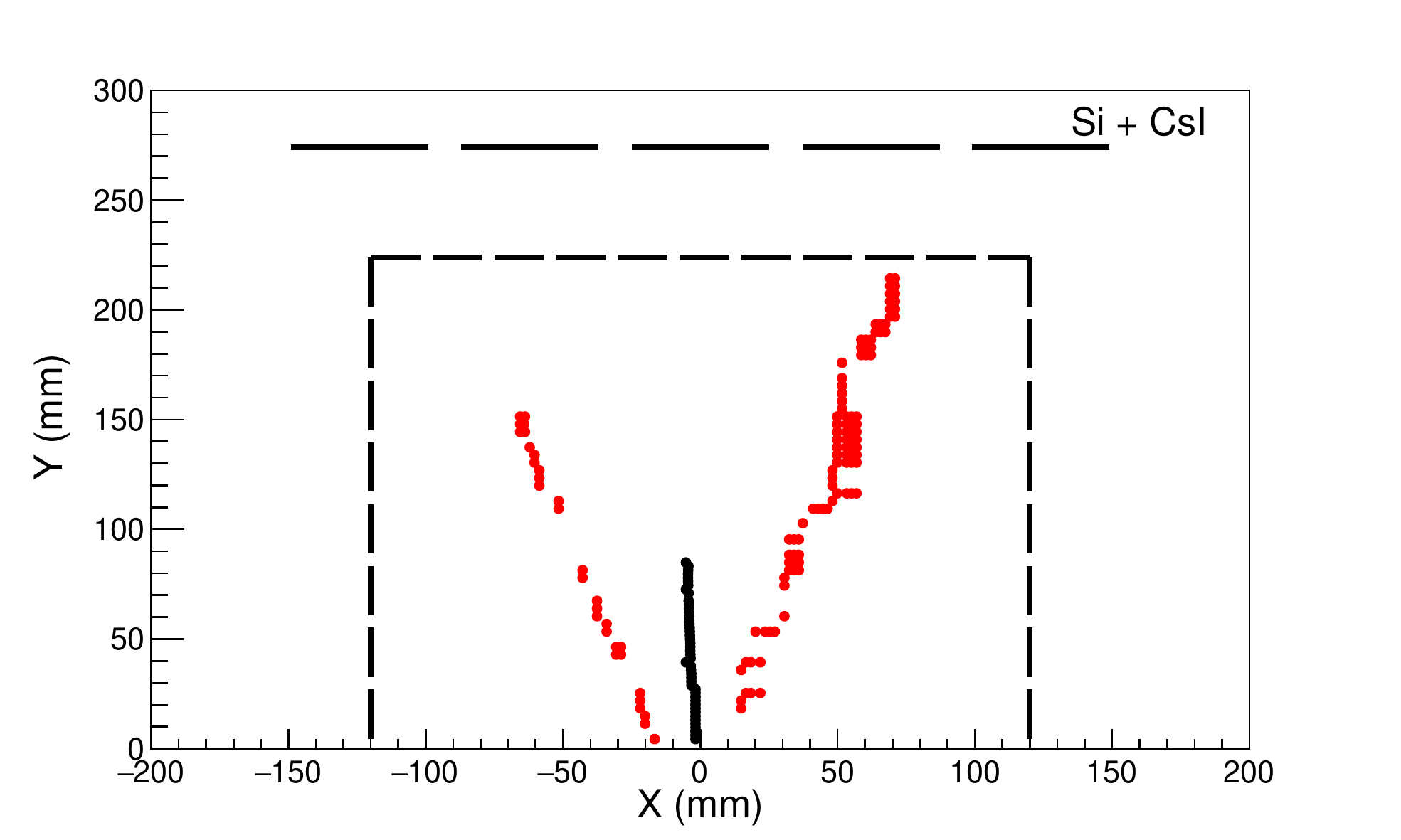}
\caption{(Color online) A 2D projection onto the micromegas plane for a typical $^8$B+p inelastic scattering event, with subsequent proton decay of a $^8$B excited state. The $^7$Be recoil track is shown in black. The proton tracks, produced by matching strips and chains in the multiplexed high-gain side region of the micromegas detector, are shown in red. Active area of the micromegas detector is shown by the dashed lines and the location of the Si+CsI(Tl) telescopes is shown with bold solid lines. For this event the reaction vertex is located outside of the active region of micromegas (negative Y-values). It is not shown but it can be easily reconstructed. The proton to the right of the $^7$Be recoil track produced a trigger in a Si detector. The proton to the left of the $^7$Be recoil track did not make it to the Si array. \label{fig:InelasticEvent}}
\end{figure}

\begin{figure}
  \centering
  \includegraphics[scale=0.45]{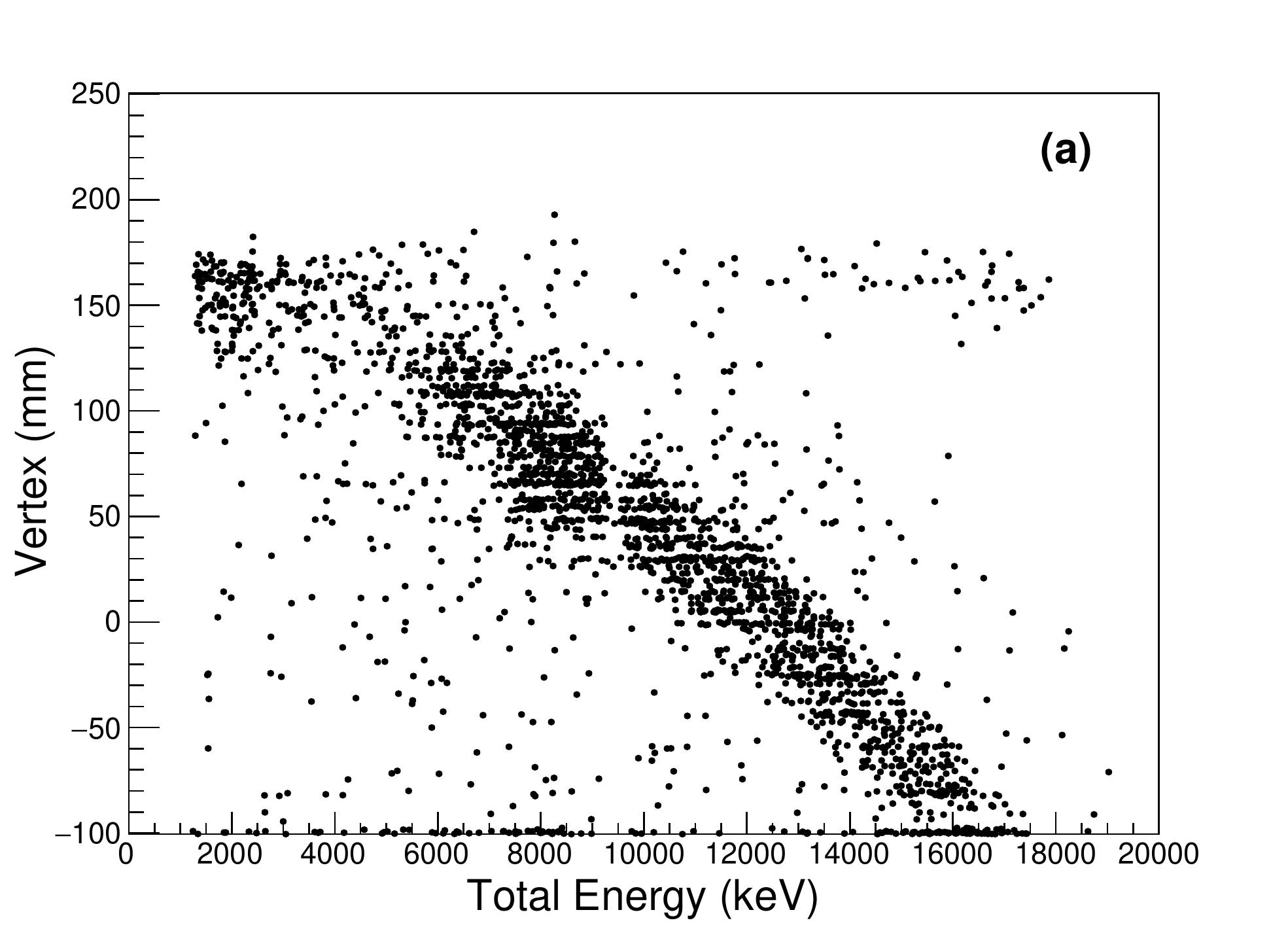}
  \includegraphics[scale=0.45]{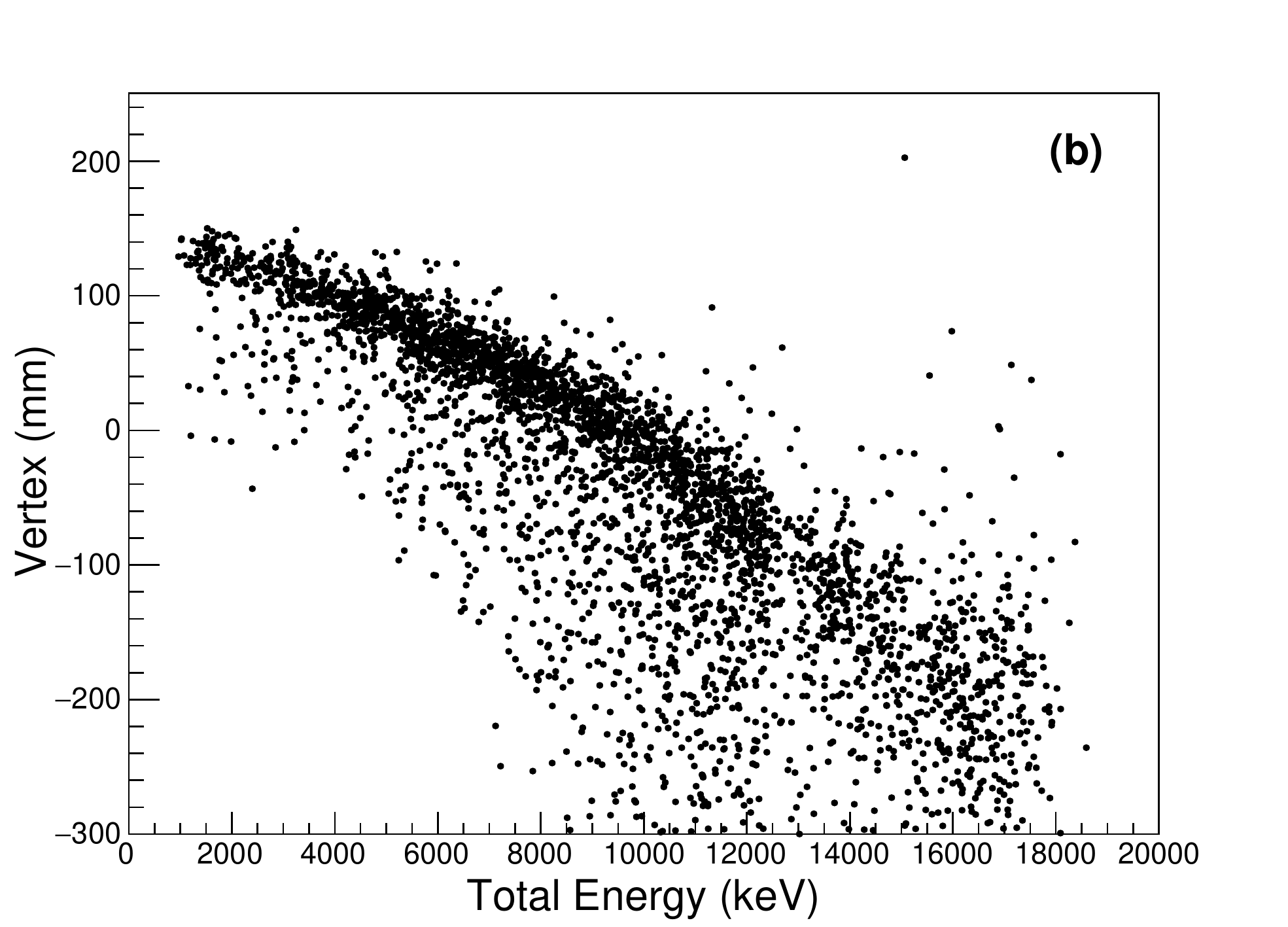}
  \caption{Vertex location plotted against the proton energy as measured by Si and CsI(Tl) detectors. Vertex location is measured relative to the start of the Micromegas detector (0 mm) where positive vertex positions are further downstream towards the forward Si detectors and negative vertex positions occur further upsteam of the Micromegas detector, outside of the active region of the TPC. The top panel (a) corresponds to the events that produce a proton track in the central region of the micromegas only, and the bottom panel (b) are the events that produce a proton track in the side regions.   \label{fig:vertexTotalEnergy}}
\end{figure}

\begin{figure}
  \centering
  \includegraphics[scale=0.45]{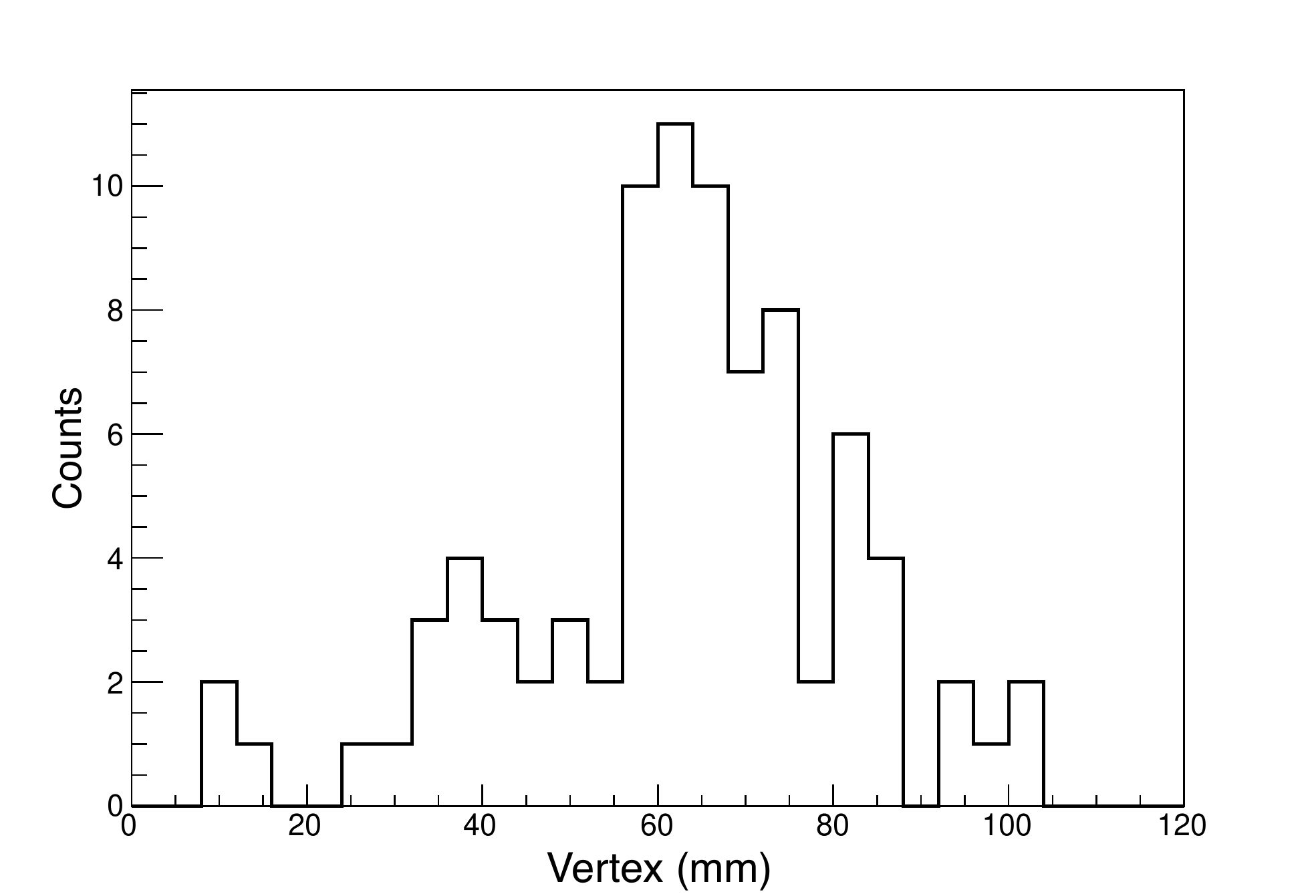}
  \caption{The vertex location distribution for events with a c.m.\ energy between 2.4 - 2.5 MeV for events in the detectors located at c.m.\ angles 100-145$^{\circ}$. The vertex is relative to the start of the Micromegas detector (0 mm) where positive vertex positions are further downstream. \label{fig:vertexCMEnergy24_25}}
\end{figure}

The inelastic p+$^8$B scattering have also been observed. There are no proton-bound excited states in $^8$B. Therefore, any p+$^8$B inelastic scattering event will produce two protons and a $^{7}$Be recoil. An example of such event is shown in Fig. \ref{fig:InelasticEvent}. These events are excluded from the analysis presented in this paper.

\section{\label{sec:Results} Results}

\begin{figure}
\includegraphics[scale=0.65]{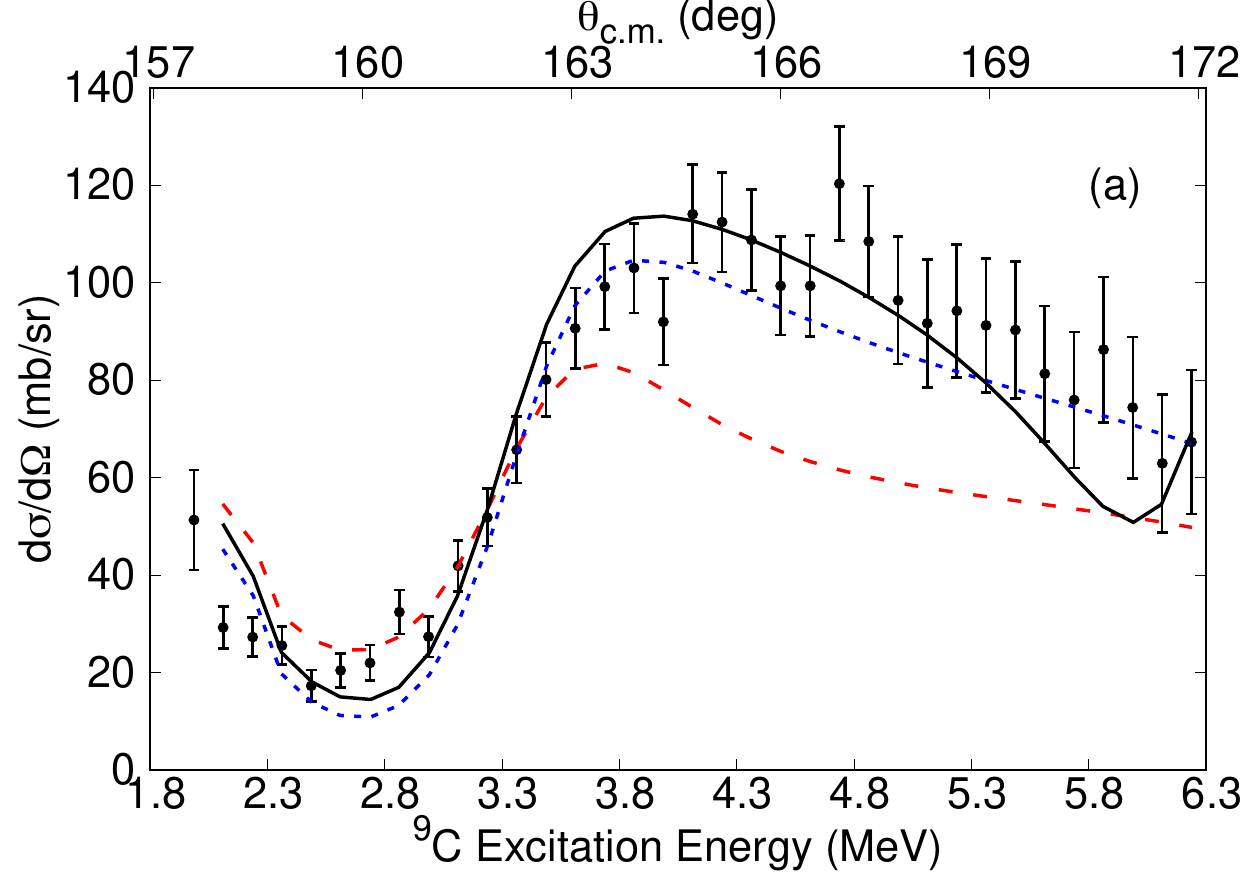}
\includegraphics[scale=0.65]{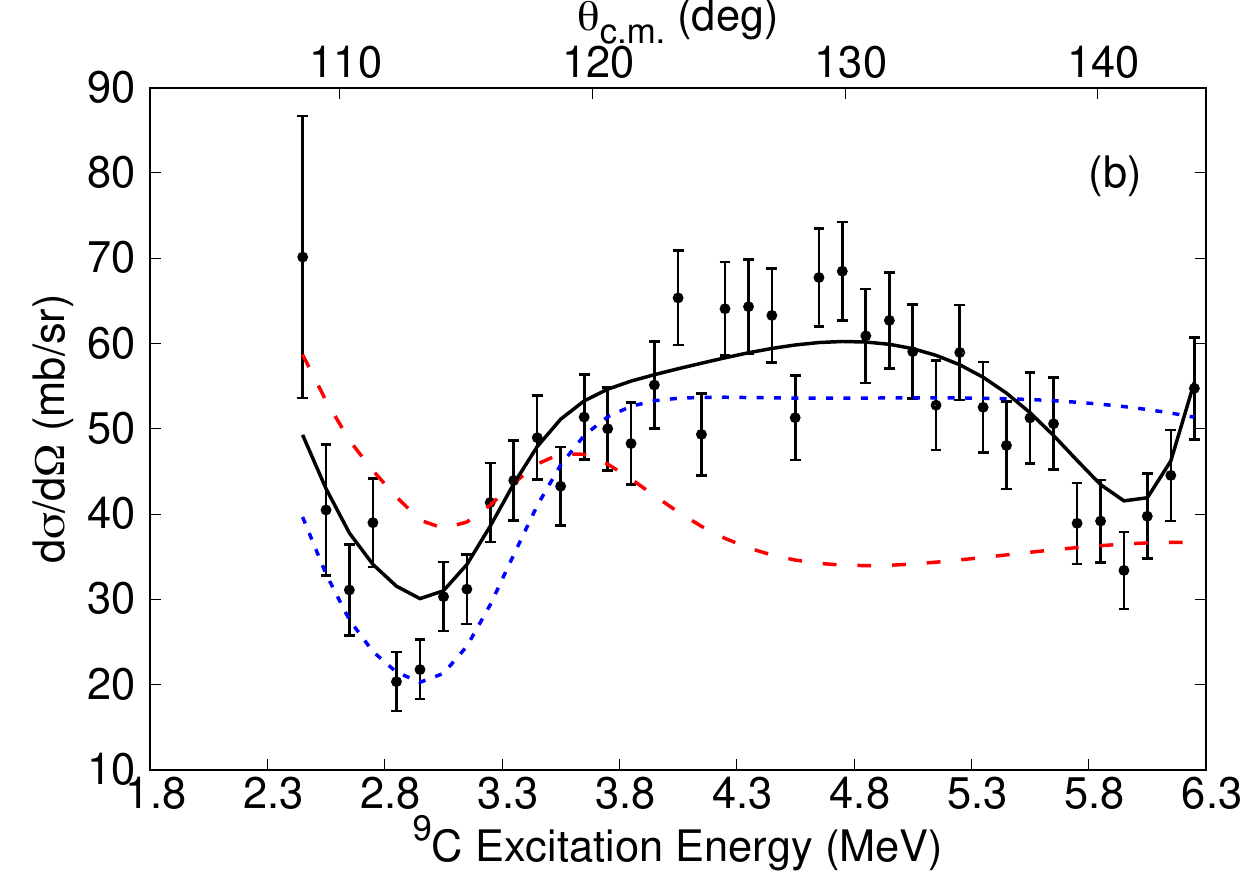}
\caption{(Color online) Excitation function for $^8$B+p elastic scattering in the angular range of 157 - 172 $^{\circ}$ (a) and 100 - 145$^{\circ}$ (b) in c.m.\ The red dashed curve is the R-matrix calculations using the phase shifts shown in Fig.\ \ref{fig:phases}. The blue short-dashed curve includes the 1/2$^-$, 5/2$^-$, and 5/2$^+$ states while the black solid curves includes a 7/2$^{-}$ on top of the 1/2$^-$, 5/2$^-$, and 5/2$^+$ states. \label{fig:EF}}
\end{figure}

The excitation functions for $^8$B+p elastic scattering for two angular ranges are shown in Fig.\ \ref{fig:EF}. The scattering angle for the detected proton recoils is a function of energy - the smaller the energy the larger the scattering angle in the lab. frame, and therefore the smaller it is in the c.m.\ for detected proton recoils. Note that zero proton scattering angle in the lab frame corresponds to 180$^{\circ}$ in c.m.\ frame in inverse kinematics. We will refer to the ``direct kinematics'' c.m.\ scattering angle throughout this paper. The absolute normalization of the cross section was performed by summing the total number of $^{8}$B ions measured in the IC, and taking into account the actual solid angle as determined by the average vertex location for each energy bin. The only systematic uncertainty in absolute normalization is related to the uncertainty of the effective target thickness per energy bin, which was calculated using specific energy losses given by code SRIM \cite{Srim:2010} and considered to be negligible as compared to the statistical uncertainty. Note that the shape of the $^8$B+p excitation function around 164$^{\circ}$ scattering angle (Fig. \ref{fig:EF}a) is very similar to the results of the previous study (Ref. \cite{Rogachev:2006}, Fig. 7), but the absolute normalization is different by about 10-15\%. This is not surprising, given that various backgrounds had to be subtracted in \cite{Rogachev:2006} and that the number of accumulated $^8$B ions was not counted directly, but evaluated using Faraday cup for the primary beam and an assumption that $^8$B/$^6$Li ratio remains constant.

R-matrix analysis of the excitation functions was performed using the code MinRmatrix \cite{Johnson:2008}. Only two channels were included explicitly in the analysis, the elastic scattering and the inelastic scattering populating the first excited state of $^8$B, the 1$^+$ at 0.77 MeV. A channel radius of 4.5 fm was used for both of these channels. First, we tried to reproduce the observed excitation functions using only the known states in $^9$C - the 3/2$^-$ ground state, 1/2$^-$ at 2.2 MeV, and the 5/2$^-$ at 3.6 MeV, as in \cite{Rogachev:2006}. It is typical to include the ``background'' resonances at high energy in the R-matrix calculations to emulate the contribution from the higher lying resonances that are not taken into account explicitly. These ``background'' resonances are normally considered free parameters. In the attempt to reduce the number of free parameters and to make the R-matrix analysis as realistic as possible we have adopted a different approach in this paper. The 1/2$^-$, 3/2$^-$, and 5/2$^-$ p-waves were constrained by solving the Schr\"odinger equation for a single proton in the field of $^8$B(g.s.). A Woods-Saxon shape for the p+$^8$B interaction potential was adopted, with diffuseness set to 0.65 fm and the reduced radius set to r=1.2 fm (R=1.2$\sqrt[3]{8}$ fm). The potential well depth was adjusted to fit the energies of the known states in $^9$C with respect to the proton decay threshold. For example, the 3/2$^-$ ground state is bound by 1.3 MeV and the potential depth required to reproduce this binding energy is 50.45 MeV. The Coulomb interaction was approximated by a potential of a uniformly-charged sphere with reduced radius of $r_{c}$=1.3 fm. The parameters of the high energy ``background'' resonances in R-matrix calculations were then tuned for each partial wave so that the resulting R-matrix phase shift reproduces that of the Schr\"odinger equation in the energy range relevant for this analysis (from 1 to 5 MeV in c.m.), as shown in Fig.\ \ref{fig:phases}. A perfect match is achieved for the 3/2$^-$ and 5/2$^-$ partial waves. The 1/2$^-$ R-matrix phase shift deviates from the solution of the single-particle Schr\"odinger equation because its width is about a factor of two smaller than a single-particle width, if we adopt the most recent experimental value ($\Gamma$= 52$\pm$11 keV) \cite{Brown:2017}. Since the 1/2$^-$ state is below the energy range measured in this work, the overall influence of the 1/2$^-$ partial wave on the excitation function is minimal and the specific choice made above has no influence on the final result.

Further reduction of the number of free parameters in the R-matrix fit was achieved by noticing that each of the $\ell=1$ partial waves is dominated by one of the possible channel spins (S=3/2 or S=5/2) in the entrance channel. Naturally, the 1/2$^-$ and 7/2$^-$ p-shell states can only be populated with channels spins 3/2 and 5/2 respectively (if we exclude the $\ell=3$ contribution). The 3/2$^-$ ground state is known to be dominated by a proton in the 1p3/2 shell, and the 5/2$^-$ state corresponds to a proton in the 1p1/2 shell \cite{Wuosmaa2005,Rogachev:2006}. Re-coupling from JJ coupling scheme to the LS coupling leads to the dominant contribution of the S=3/2 channel spin for the 3/2$^-$, and the S=5/2 channel spin for the 5/2$^-$. Therefore, we have restricted the R-matrix calculation to channel spin 3/2 for the 1/2$^-$ and 3/2$^-$ partial waves, and to channel spin 5/2 for the 5/2$^-$ and 7/2$^-$ partial waves.

This way the excitation function calculated using R-matrix approach and shown in Fig.\ \ref{fig:EF} by the dashed red curve has no free parameters.

While the shape of the 155$^{\circ}$ - 170$^{\circ}$ angular range excitation function is described, the absolute magnitude of the cross section is underestimated. Most importantly, the shape of the excitation function for the 105$^{\circ}$ - 145$^{\circ}$ angular range is wrong. Adding the tentative 3/2$^-$ at 4.1 MeV (as it was done in \cite{Rogachev:2006}) improves the fit at large c.m.\ scattering angle but does not help to reproduce the 105$^{\circ}$ - 145$^{\circ}$ angular range. We have found that fitting the deep minimum observed in the $^8$B+p excitation functions at 1.5 MeV for smaller c.m.\ scattering angles (around 110$^{\circ}$) requires a strong destructive interference between the $\ell$=0 partial wave and Coulomb amplitude. A broad $J^{\pi} = 5/2^{+}$ state, located at an excitation energy around 4.3 MeV, achieves the desired effect at smaller c.m.\ scattering angles, and also improves the fit at large c.m.\ scattering angles. The R-matrix calculation that includes the 1/2$^-$, 5/2$^-$, and 5/2$^+$ states is shown in Fig.\ \ref{fig:EF} with the short-dashed blue curve. Other spin-parity assignments for the new state were considered. The 1/2$^+$ cannot decay to the $^8$B(2$^+$) ground state with $\ell=0$ (only $\ell=2$) and therefore does not produce the required interference pattern. The 3/2$^+$ spin-parity assignment was tried, but it resulted in substantially worse agreement. Further improvements are achieved by introducing an $\ell=1$ $J^{\pi} = 7/2^{-}$ state with an excitation energy of 6.4 MeV (5.1 MeV in c.m.) and a width of 1.1 MeV as shown in Fig.\ \ref{fig:EF} with the solid black curve. Although our spectrum does not extend beyond 5.0 MeV and therefore we cannot make definitive conclusions, a $7/2^{-}$ was suggested by the Continuum Shell Model calculations at 6.3 MeV \cite{Volya:2014}.  Also, there is a state in the mirror nucleus, $^9$Li, at similar excitation energy - 6.43 MeV, the spin-parity assignment for which is unknown. We included the 7/2$^{-}$ at 6.4 MeV into our final fit, but consider it tentative.

\begin{figure}
\includegraphics[scale=0.65]{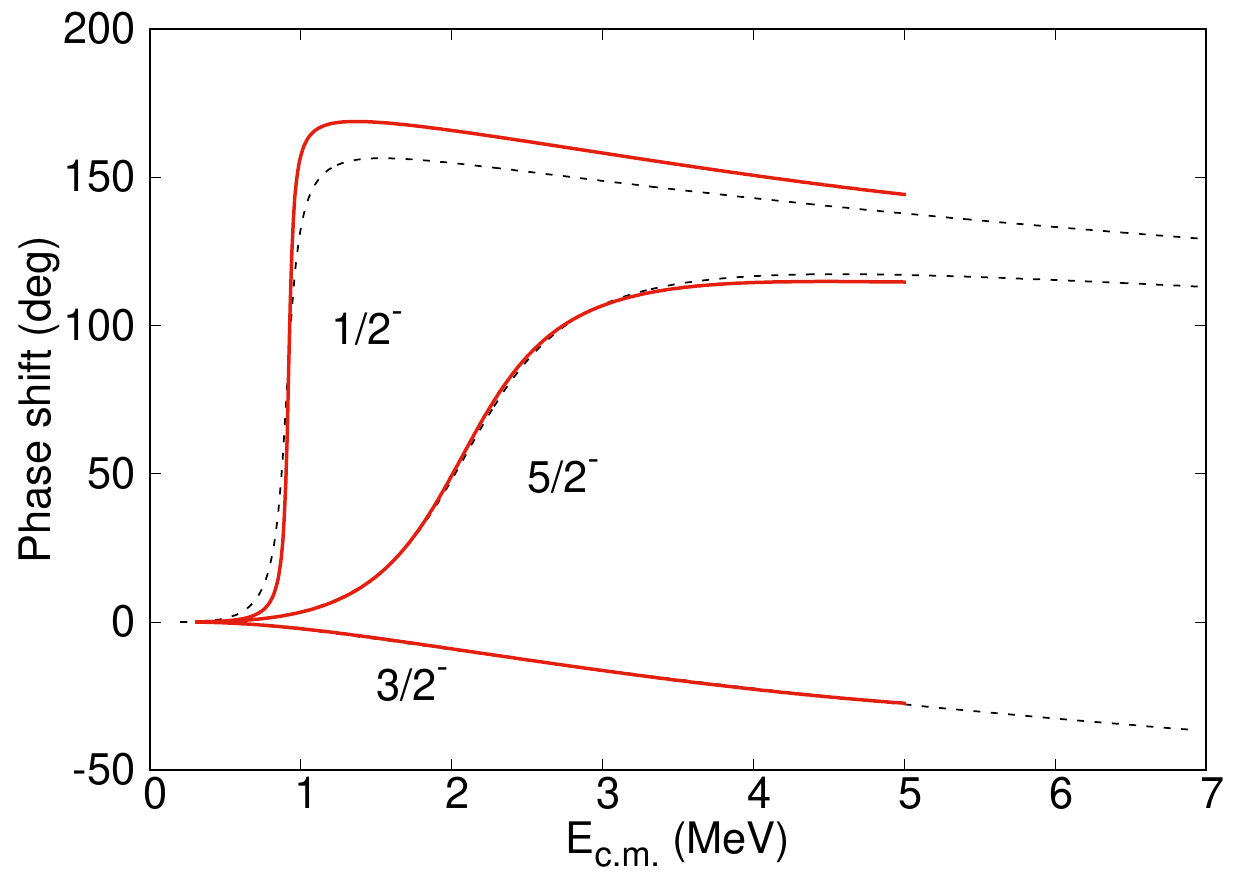}
\caption{ (Color online) The dashed black curves are the phase shifts for the $\frac{3}{2}^-$ and $\frac{1}{2}^-$ partial waves calculated using Woods-Saxon potentials that reproduce the binding energy of the $^9$C $\frac{3}{2}^-$ ground state (1.3 MeV) and the c.m.\ energy of the $\frac{1}{2}^-$ first excited state (0.92 MeV) \label{fig:phases}. The R-matrix phase shifts are shown as red solid curves. \label{fig:phases}}
\end{figure}

Due to the nature of this excitation function that is defined by broad overlapping resonances - we used additional {\it a priori} constrains to reduce the number of free parameters in the fitting procedure. The excitation energy of the 5/2$^-$ state is well established and is in good agreement between the Refs. \cite{Rogachev:2006} and \cite{Brown:2017}. In particular, in Ref.\ \cite{Brown:2017} the 5/2$^-$ resonance energy is well defined (but not necessarily the width, see comments in section \ref{sec:Discussion}), so the excitation energy of this state was fix at 3.6 MeV (within the uncertainties given in \cite{Brown:2017}). We have studied the quality of the fit as a function of the widths of the 5/2$^-$ state by manually varying it around the best fit value and fitting the parameters of the 5/2$^+$ and 7/2$^-$ states for each manual iteration. The best fit is achieved for the 5/2$^-$ width of 1.1 MeV with one standard deviation of 300 keV. The sensitivity of the fit to the parameters of the 5/2$^+$ state was evaluated by fixing the 5/2$^-$ at its best fit values and manually varying the excitation energy and widths for the 5/2$^+$. The result for the 5/2$^+$ is E$^*$ = 4.3$\pm$0.3 MeV and $\Gamma$=4.0$^{+2.0}_{-1.4}$ MeV.

\section{\label{sec:Discussion} Discussion}

The excitation function for $^8$B+p elastic scattering at large c.m.\ angles is dominated by the 5/2$^-$ state, confirming the results of Ref.\ \cite{Rogachev:2006}. There are two main advantages of the presented data over that of Ref.\ \cite{Rogachev:2006}: (1) no background subtraction was necessary, and (2) the much wider range of scattering angles is measured in the present active target experiment. The latter provides clear evidence for a broad $\ell=0$ 5/2$^+$ state, which plays a dominant role at smaller c.m.\ scattering angles (close to 90 $^{\circ}$). We confirm the previous findings of Ref.\ \cite{Rogachev:2006} that there is no evidence for an excited state at 3.3 MeV reported in \cite{Golovkov:1991}. The 5/2$^-$ state has also been recently observed in the inelastic scattering of a $^{9}$C beam on a $^{9}$Be target \cite{Brown:2017}. The best fit total width for this state is notably different in this work and in \cite{Rogachev:2006}, as compared to \cite{Brown:2017}. It is 630 keV in \cite{Brown:2017} and 1.1 MeV in the present work. It is possible that the background subtraction procedure applied in \cite{Brown:2017} is a cause for this discrepancy. The background constitutes 2/5 of the observed yield in Ref.\ \cite{Brown:2017} at the 5/2$^-$ resonance maximum, while there is no background in the excitation function for $^8$B+p elastic scattering obtained in this work. The observed 1.1 MeV width of the 5/2$^-$ state corresponds to the single particle width - the p+$^8$B(g.s.) spectroscopic factor (SF) for this state is 0.8+/-0.2. This is in agreement with the SF of 0.93(20) measured for the mirror 5/2$^-$ state in $^9$Li \cite{Wuosmaa2005} using $^8$Li(d,p) reaction and also with predictions of the {\it ab initio} models \cite{Wuosmaa2005}. The result of Ref.\ \cite{Brown:2017} would infer the SF of 0.45 for this state - in disagreement with {\it ab initio} calculations and experimental data on SF in $^9$Li. For completeness we note that the total width of the 5/2$^-$ state at 4.3 MeV in $^9$Li was measured in three different experiments and the results are not consistent. It is 250(30) keV in \cite{Young1971}, 100(30) keV in \cite{Ajze1978}, and 60(45) keV in  \cite{Heilbronn1989}. The single particle width for this state in $^9$Li is 180 keV, so while the former value would be more in agreement with the results of this work as well as the {\it ab initio} predictions \cite{Wuosmaa2005,Nollett2012} and the results of the $^8$Li(d,p) measurements  \cite{Wuosmaa2005}, the latter would be more in-line with the Ref.\ \cite{Brown:2017}.

A broad ($\Gamma$=2.75(11) MeV) excited state has been observed at 4.4 MeV excitation energy in Ref.\ \cite{Brown:2017}. This state was assigned a positive parity in \cite{Brown:2017}, but reasons for that assignment were not discussed. We have observed clear evidence for a broad $\ell=0$ 5/2$^+$ ($\Gamma$=4.0$^{+2.0}_{-1.4}$ MeV)  state at 4.3 MeV, which is in good agreement with \cite{Brown:2017}.

\begin{figure}
  \centering
  \includegraphics[scale=0.65]{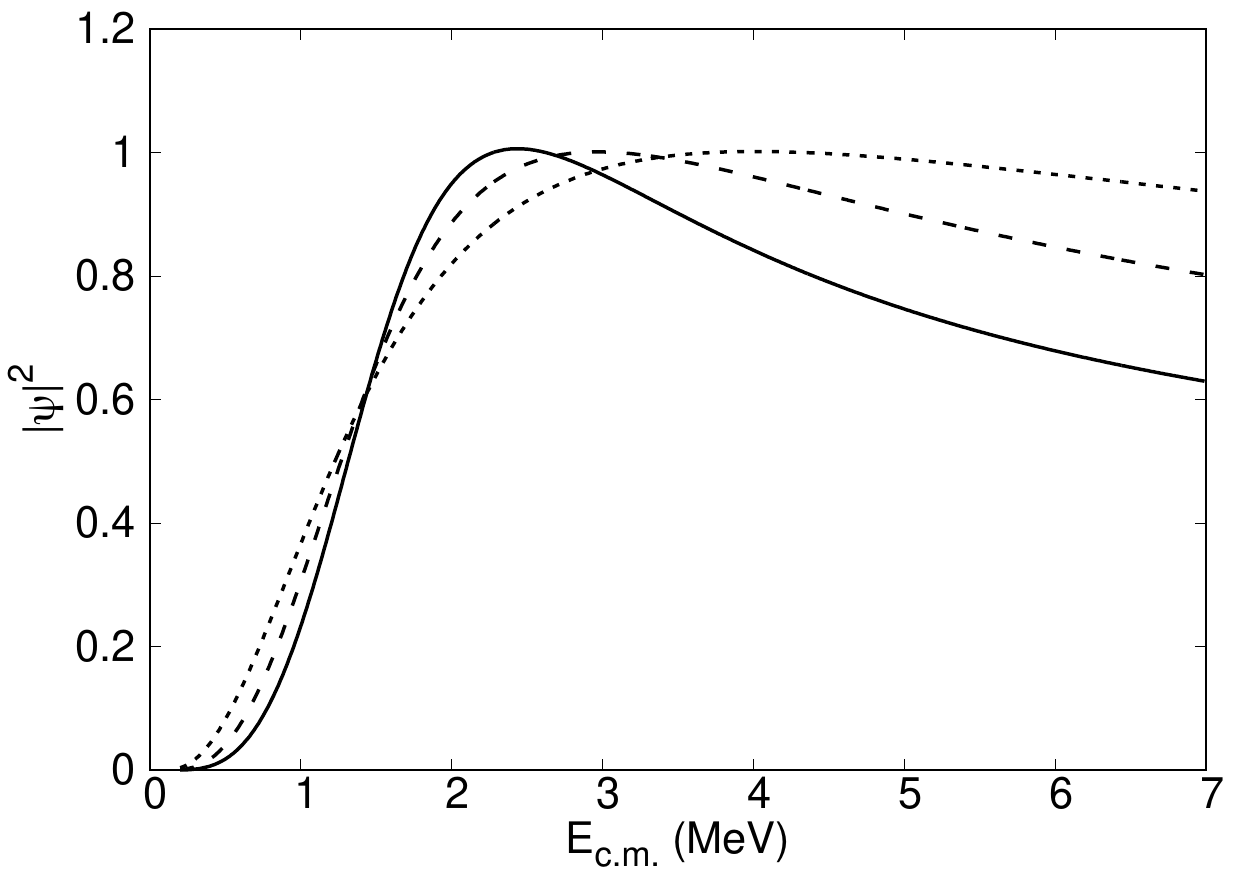}
  \caption{Square of the amplitude of the proton wave function (arb. un.) at 1.00 fm for the s-wave states in $^8$B (short-dashed curve), $^9$C (dashed curve), and $^{10}$N (solid curve) calculated using the Woods-Saxon potential with V=-58.0 MeV, R=1.2$\sqrt[3]{A}$ fm, a=0.65 fm, and the Coulomb potential due to the uniformly-charged sphere of radius 1.3$\sqrt[3]{A}$, where A= 7, 8 and 9 for the $^8$B, $^9$C, and $^{10}$N, respectively. \label{fig:swaves}}
\end{figure}

This result firmly establishes the onset of the 2s1/2 shell in T=3/2 A=9 and gives us an opportunity to discuss systematics for the 2s-shell in light proton rich p-shell nuclei $^8$B, $^9$C, and $^{10}$N. To do that we employ the potential model mentioned above. We fit the potential depth to reproduce the 5/2$^+$ resonance energy in $^9$C. Since the corresponding phase shift never reaches 90$^{\circ}$, we investigate the resonance behavior by plotting the square of the amplitude of the wave function at certain fixed distance from the origin (1.0 fm) as a function of c.m.\ energy \cite{Mukhamedzhanov2010}. The dashed curve in Fig.\ \ref{fig:swaves} corresponds to the single-particle 5/2$^+$ in $^9$C at 4.3 MeV (3.0 MeV in c.m.), as observed experimentally. We can now explore the s-wave in $^8$B and $^{10}$N by making an assumption that the s-wave proton-core interaction is the same for all three nuclei ($^8$B, $^9$C, and $^{10}$N). So, keeping parameters of the Woods-Saxon potential exactly the same (only the reduced mass and charge are different) we get the 2s resonances in all three nuclei, as shown in Fig.\ \ref{fig:swaves}. This result is in remarkable agreement with the experimental data. The 2s-shell is located at c.m.\ energy of 2.3$\pm$0.2 MeV in $^{10}$N \cite{Hooker2017}. This is where a maximum of the amplitude of the wave function for the $^{10}$N is predicted by the potential model (Fig.\ \ref{fig:swaves}). For the $^7$Be+p the most recent comprehensive R-matrix analysis \cite{Brune2019} gives -3.18$^{+0.55}_{-0.50}$ fm scattering length for the $\ell=0$ 2$^-$ partial wave. The potential model $\ell=0$ phase shift for the $^7$Be+p scattering calculated using parameters given in the caption for Fig.\ \ref{fig:swaves} corresponds to the scattering length of -5.5 fm, in fair agreement with the actual experimental value, especially considering the simplicity of the underlying approach that completely ignores the possible isospin dependence of the nucleon-core interaction potential. It is also in perfect agreement with the {\it ab initio} calculation by P. Navratil, et al., \cite{Navratil2011}, which predicts -5.2 fm.

\section{\label{sec:Conclusion} Conclusion}

We have studied the structure of $^{9}$C by measuring an excitation function for $^{8}$B+p elastic scattering in a broad energy and angular range using the active target approach. This experiment was a commissioning run for Texas Active Target detector system (TexAT). In addition to the two previously known negative parity excited states in $^{9}$C, $J^{\pi} = 1/2^{-}$ and $5/2^{-}$, a broad positive parity state $J^{\pi} = 5/2^{+}$ has been observed at around 4 MeV. This state has a single-particle nature and therefore we have an opportunity to experimentally determine the location of the 2s-shell in the A=9 T=3/2 system for the first time. It would be interesting to compare this result with predictions of the {\it ab initio} models, but these are not available at the moment. Since the state is very broad, the meaningful comparison with theory can only be made if continuum is consistently taken into account. Therefore, it is not a surprise that the bound-state AMD calculations \cite{Kanada2001} significantly overestimate the 2s shell energy in $^9$C - by a factor of two. In addition to the $J^{\pi} = 5/2^{+}$ state at 4 MeV, there is some evidence for a $J^{\pi} = 7/2^{-}$ state at 6.4 MeV with a width of 1.3 MeV. Although we cannot make definitive conclusions regarding this state since the excitation function has not been measured to high enough excitation energies and we've only observed the low energy tail of this resonance, it does agree with the prediction of the CSM \cite{Volya:2014} for the 7/2$^-$ excitation energy. The $J^{\pi} = 5/2^{+}$ observed in this measurement is the first conclusive observation of any sd-shell state in $^{9}$C or any other member of A=9 T=3/2 isospin multiplet. This observation was made possible by application of an active target approach that allowed to measure the $^8$B+p excitation functions in wide range of scattering angles.

We thank Rui de Oliviera \& Bertrand Mehl at CERN for their technical assistance in developing the Micromegas pad plane. This work was supported by the U.S. Department of Energy, Office of Science under grant number \#DE-FG02-93ER40773, by the National Nuclear Security Administration through the Center for Excellence in Nuclear Training and University Based Research (CENTAUR) under grant number \#DE-NA0003841, and by the Nuclear Solutions Institute at Texas A\&M University.

\bibliography{9C}{}

\end{document}